\documentstyle[12pt,preprint,aps,prl]{revtex}
\begin{document}
\draft
\tighten

\title{LIMITS ON THE DOMAIN OF COUPLING CONSTANTS\\
 FOR BINDING \protect{$N$}-BODY SYSTEMS\\
 WITH NO BOUND SUBSYSTEMS
\footnote{Previous title: Limits on the Window for Halo Phenomena}
\footnote{to be published in Phys.\  Rev.\  Lett.}
}
\author{Jean-Marc Richard$^{(1),(2)}$ and Sonia Fleck$^{(3)}$}
\address{$^{(1)}$Institut des Sciences Nucl\'eaires\\
53, avenue des Martyrs, 38026 Grenoble, France\\
$^{(2)}$ European Centre for Theoretical Studies
in Nuclear Physics and Related Areas (ECT*)\\
Villa Tambosi, via delle Tabarelle 286, I--38050 Villazzano (Trento),
Italy\\
$^{(3)}$Institut de Physique Nucl\'eaire\\
43, boulevard du 11 Novembre 1918,
 69622 Villeurbanne, France}
\date{\today}
\maketitle

\begin{abstract}
We study the domain of coupling constants
for which a 3-body or 4-body system is bound while none of its
subsystems is bound. Limits on the size of the domain are derived
from a variant of the Hall--Post inequalities which relate $N$-body
to
$(N-1)$-body energies at given coupling. Possible applications to
halo nuclei
and hypernuclei are briefly outlined.
\end{abstract}


For the sake of clarity, we shall define in this Letter a ``halo''
as a 3-body quantum system  that is bound
while none of its 2-body subsystems has a discrete spectrum. More
generally an
$N$-body halo is bound while
none of its subsets is stable against spontaneous dissociation. This
is more
restrictive than the usual meaning of a weakly bound system with a
very
extended
wave function. These systems are sometimes called ``Borromean''
\cite{Bang},
after
Borromean rings, which are interlaced in such a subtle topological
way that if
any of them is removed,  the other two would be unlocked.

Halo states
are seen in nuclear physics \cite{Bang,Jensen3b}. For instance the
$(\alpha,n,n)$ system is bound ($^6$He), while  $(\alpha,n)$ and
$(n,n)$
systems are both  unbound.
There is a cooperative effort of all attractive potentials to achieve
the
binding of $^6$He.

The halo phenomenon shows up in simple potential models,
as seen from explicit calculations on specific isotopes
\cite{Bang,Jensen3b,Chinois}, or from the rich literature on the
related
Efimov effect \cite{Efimov} or Thomas  collapse \cite{Thomas},
which prove that a 3-body system is, indeed, more easily bound than a
2-body
one.

Consider a short-range potential $gV(r)$
acting between two particles of mass $m$ separated by $r$.
Even if $V(r)$ is attractive,  a
minimal strength $g_2^c/m$ is needed to achieve binding, where
$g_2^c$ is
independent of $m$.
For instance, in a Yukawa potential $V=-(\exp(-\mu r))/r$, one can
fix
the energy and distance scales so that $m=\mu=1$ without loss of
generality,
and one finds
$g_2^c\simeq 1.680$ \cite{Jackson}.
(A simple argument by Dyson and Lenard \cite{Dyson} shows that
$g_2^c>\sqrt2$.)
Now, if one considers three identical bosons with mass $m=1$
interacting
through $\sum(\exp-r_{ij})/r_{ij}$, where $r_{ij}$ is the relative
distance
between particles $i$ and $j$, one can look at bound states by
variational
methods or by solving the Faddeev equations, and one finds binding
for $g\ge
g_3^c$, with
$g_3^c/g_2^c\simeq 0.804$. We call this a $20\%$ window for halo
binding.

If one repeats the above calculations for other short-range
potentials, one
finds similar values for $g_3^c/g_2^c$, for instance 0.801 for an
exponential
form
$V(r)=-\exp(-r)$, or 0.794 for a Gaussian $V=-\exp(-r^2)$. Such
quasi-universality, already noticed in \cite{Gillepsie}, is not too
surprising.
In the weak coupling limit, the wave function extends very
far away and does not really probe the detailed structure of the
potential.
Any short-range interaction is almost equivalent to a delta function
in
this limit.

It seems therefore very likely that the window for halo phenomena is
limited,
i.e.,  $g_3^c/g_2^c$ cannot be made arbitrarily small by tuning the
shape of
$V(r)$. We shall show below that
\begin{equation}\label{g3vsg2}
{g_3^c\over g_2^c}\ge {2\over3}
\end{equation}
for any potential, and we outline how to derive similar inequalities
for
asymmetric 3-body systems, or for 4-body or more complicated systems.

The method of deriving (\ref{g3vsg2}) is directly inspired by the
Hall--Post
inequalities \cite{Post}, which have been recently rediscovered
\cite{BMR1},
applied to hadron spectroscopy \cite{BMR2} and generalized to the
case of
unequal masses \cite{BMR3}.
Earlier applications were mostly motivated by considerations on the
stability
or instability of matter. The Hamiltonian for three bosons of mass
$m$ can be
written as
\begin{eqnarray}\label{H3-decomp}
H_3&=&\sum_{i=1}^{3}{
{\bf p}_i^2\over2m}+g\sum_{i<j}V(r_{ij})\nonumber\\
&=&{\left(\sum{\bf p}_i\right)^2\over6m}
+\sum_{i<j}{2\over3m}\left({\bf p}_i-{\bf
p}_j\over2\right)^2+gV(r_{ij}),
\end{eqnarray}
i.e., introducing the translation-invariant part
 $\,\widetilde{\!H}_N$  of the Hamiltonian $H_N$
\begin{equation}
\label{H3toH2}
\,\widetilde{\!H}_3(m,g)=\sum_{i<j}\,\widetilde{\!H}_2^{(ij)}(3m/2,g)=
{2\over3}
\sum_{i<j}\,\widetilde{\!H}_2^{(ij)}(m,3g/2).
\end{equation}
Hence, from a simple application of the variational principle,
 the ground-state energies $E_N$ for a given (large enough) coupling
$g$
satisfy
\begin{equation}\label{HallPost32}
E_3(m,g)\ge 3E_2(3m/2,g)=2E_2(m,3g/2),
\end{equation}
which is the simplest form of the Hall--Post inequalities.
The decomposition (\ref{H3toH2}) also implies that if
$\,\widetilde{\!H}_3$
has to support a bound state, each $\,\widetilde{\!H}_2$ should
produce a negative expectation value in the corresponding wave
function, and
thus
$3g/2\ge g_2^c$, q.e.d.

A straightforward generalization of (\ref{H3toH2}) to $N$ bosons is
\cite{BMR1}
\begin{equation}
\label{HNvsHN-1}
\widetilde{\! H}_N(m,g)={1 \over N-2}\sum_{k=1}^{N}\,\widetilde{\!
H}_{N-1}^{[k]}
\left( {Nm\over N-1},g \right)
,\end{equation}
where the superscript in $\,\widetilde{\! H}_{N-1}^{[k]}$ means that
the $k$-th particle is omitted. Saturating with the ground state
of $\,\widetilde{\!H}_N$
gives
\begin{equation}
\label{gNvsgN-1}
g_N^c\ge{N-1\over N}g_{N-1}^c,
\end{equation}
i.e. $Ng_N^c$ increases with $N$.

For numerical illustration in the $N=4$ case, we adopted a
variational
method that is widely used in quantum chemistry \cite{quantumc}.
It is based on trial wave functions of the type
\begin{equation}
\label{trial}
\Psi({\bf x}_i)=\sum_n
c^{(n)}\exp\Bigr[-{1\over2}\sum_{i,j}a_{ij}^{(n)}
{\bf x}_i\cdot {\bf x}_j\Bigr],
\end{equation}
where $\{{\bf x}_i\}$ is a set of relative Jacobi coordinates.
Symmetry is
properly
implemented by imposing relations
between neighbouring coefficients $c^{(n)}$ and definite-positive
matrices $a^{(n)}$. After numerical minimization, we
obtain $g_4^c/ g_2^c \simeq 0.67$
for a Yukawa potential, i.e.\ a $13\%$ window for a genuine 4-boson
halo, once $g_4^c/g_2^c$ is subtracted from $g_3^c/g_2^c$.

Note that the bounds (\ref{g3vsg2}) and (\ref{gNvsgN-1}) are
not expected to be saturated, since the Hall--Post inequalities
become equalities only for harmonic oscillators, which are
far from the short-range potentials we consider here.

Similar inequalities can be written down in a variety of situations.
Let us give some examples. Consider first $N$  identical particles
with mass $m=1$ in the field of an infinitely massive source.
The Hamiltonian is defined as
\begin{equation}
\label{Hamiltonian-more-gen}
H_N=\sum_{i=1}^{N}
\left[ { {\bf p}_i^2\over 2m}+hV(r_i) \right]
+g\sum_{i<j}W(r_{ij}).
\end{equation}
The short-range attractive potentials $V$ and $W$ can be normalized
such that a
single particle is trapped around the source for $h>1$, and two
particles
are bound together for $g>1$. The non-trivial domain is thus to be
found inside
the square $(h<1,g<1)$.

Let $h_N(g)$ be the boundary for binding $N$ particles around the
source. This
means that halo type of binding occurs between $h_N(g)$ and
$h_{N-1}(g)$.
One expects all curves to merge at A$(h=1,g=0)$ as $g\rightarrow 0$,
since the
particles become independent in this limit.
If one takes the expectation value of the identity
\begin{equation}
\label{H3vsH2-2}
H_ 3(h,g)={1\over2}\sum_{i<j}H_2^{(i,j)}(h,2g)
\end{equation}
within the ground state of $H_3$, one gets
\begin{equation}
\label{h3vsh2}
h_3(g)\ge h_2(2g),
\end{equation}
and more generally
\begin{equation}
\label{hNvshN-1}
h_N(g)>h_{N-1}\left({N-1\over N-2}g\right).
\end{equation}

For $N=2$, one can write the simple decomposition
\begin{equation}
\label{H2-decomp-1}
H_2=\left[{\alpha\over2}{\bf p}_1^2+hV(r_1)\right]+
\left[{\alpha\over2}{\bf p}_2^2+hV(r_2)\right]
+\left[{1-\alpha\over2}\left({\bf p}_1^2+{\bf
p}_2^2\right)+gW(r_{12})\right],
\end{equation}
for any $0\le\alpha\le1$. To get $\langle H_2\rangle<0$, one needs
at least one of the square brackets having a negative expectation
value.
This excludes the triangle $\{h\le\alpha,g\le(1-\alpha)\}$
shown in Fig.~\ref{Fig1}.

Two remarks on this simple lower bound are in order. First, the
actual
boundary is expected to be concave.
The couplings $h$ and $g$ enter the Hamiltonian linearly,
so if the minimum $E$ of $H_2$ vanishes at both
$P(h,g)$ and $P'$, one has \cite{Thir}
\begin{equation}
\label{concavity}
E(\lambda P+(1-\lambda) P')\ge\lambda E(P)+(1-\lambda)E(P')=0,
\end{equation}
for any $0\le\lambda\le1$.

Secondly, while the limit A$(h=1,g=0)$ of truely independent
particles
obviously belongs to the boundary,
B$(h=0,g=1)$ is likely to be in the continuum, since a weakly bound
(1,2)
system needs a minimal attraction $h$ to remain trapped by the
source.

A more elaborate decomposition leads to an improved boundary which
better
complies with the above remarks. We provisionally restore a finite
mass $M$ for
the third particle and write as in \cite{BMR3}

\begin{eqnarray}
\label{H2-decomp-2}
H_(M)&=&{{\bf p}_1^2\over2}+{{\bf p}_2^2\over2}+{{\bf p}_3^2\over2M}+
hV(r_{13})+hV(r_{23})+gW(r_{12})\nonumber\\
&=&\left({\bf p}_1+{\bf p}_2+{\bf p}_3\right)\cdot
\left(b{\bf p}_1+b{\bf p}_2+b'{\bf p}_3\right)\nonumber\\
&&+\sum_{i=1,2}\left({1+x\over1+2x}\right)^2\left({1\over2}+{1\over
M}\right)
\left({ {\bf p}_i-x{\bf p}_3\over1+x}\right)^2+hV(r_{i3})\nonumber\\
&&+{4\over(1+2x)^2}\left(x(1+x)-{1\over2M}\right)\left({{\bf p}_1-
{\bf p}_2\over2}\right)^2+gW(r_{12}),
\end{eqnarray}
where the momentum $({\bf p}_i-x{\bf p}_3)/(1+x)$ is the conjugate of
the
relative distance ${\bf r}_{i3}$, and $b$ and $b'$ are known
functions of $M$
and $x$. In the limit $M\rightarrow\infty$, we read off from
(\ref{H2-decomp-2}) that $H_2$ would
never support a bound state if
\begin{equation}
\label{H2-limits}
h\left({1+2x\over1+x}\right)^2<1\quad\text{and}
\quad g{(1+2x)^2\over4x(1+x)}<1.
\end{equation}
This is the inner part of the parabola shown in Fig.\ \ref{Fig1},
from which we
get a crude lower limit on the minimal coupling $h_3(g)$ to bind
three bosons
around the source, as per Eq.\ (\ref{h3vsh2}).

The decomposition (\ref{H2-decomp-2}) can be used for finite $M$.
Consider for
instance the case where $M=1$ and $g=0$.  One should
restrict  to $x(1+x)>1/2$ in order not to introduce a negative
reduced mass.
The two particles, which do not interact with each other, can be
bound
simultaneously
below the critical coupling $h=1$ for individual coupling. Each
particle
benefits from the increase of the reduced mass provided by the other.
However,  by choosing the optimal parameter $x$ in
(\ref{H2-decomp-2}), one can
easily deduce that the
window for halo is limited to $h>1/2+\sqrt3/4\simeq0.93$, i.e. at
most $7\%$.

In a situation where the three masses or the three couplings are
different, the
most general decomposition involves two parameters \cite{BMR3},
instead of the
single $x$ in (\ref{H2-decomp-2}). So the analysis becomes slightly
more
involved.

We now write the Hamiltonian for two identical particles of mass $m$,
and  two
others  of mass $M$
\begin{equation}
\label{H4-def}
H_4=\sum_{i=1,2}{{\bf p}_i^2\over2m}+
\sum_{j=3,4}{{\bf p}_j^2\over2M}+g_{12}V_{12}(r_{12})
+g_{34}V_{34}(r_{34})+g_{mM}\sum_{i,j}V_{mM}(r_{ij}),
\end{equation}
including up to three different potentials. It can be rewritten as
\begin{eqnarray}
\label{H4-decomp}
H_4&=&({\bf p}_1+{\bf p}_2+{\bf p}_3+{\bf p}_4)\cdot
(b{\bf p}_1+b{\bf p}_2+b'{\bf p}_3+b'{\bf p}_4)\nonumber\\
&&+a_{12}\left({{\bf p}_1-{\bf
p}_2\over2}\right)^2+g_{12}V_{12}(r_{12})
+a_{34}\left({{\bf p}_3-{\bf
p}_4\over2}\right)^2+g_{34}V_{34}(r_{34})
\nonumber\\
&&+\sum_{i,j}\bar{a}[\alpha{\bf p}_i-(1-\alpha){\bf
p}_j]^2+g_{mM}V_{mM}(r_{ij})
\end{eqnarray}
For given $\alpha$, one can solve for $b$ and  $b'$, as well as for
the inverse
masses, with the result
\begin{eqnarray}
\label{H4-inversemasses}
a_{12}&=&{1\over m}(1-\alpha^2)-{1\over M} \alpha^2,\nonumber\\
a_{34}&=&-{1\over m}(1-\alpha)^2+{1\over M} \alpha(2-\alpha),\\
\bar{a}&=&{1\over 4m}+{1\over 4M}.\nonumber
\end{eqnarray}
If one rescales the couplings to the critical value  for 2-body
binding
with the appropriate inverse reduced mass, $m^{-1}$ for $g_{12}$,
$M^{-1}$ for $g_{34}$, and $(m^{-1}+M^{-1})/2$ for $g_{mM}$, then
$H_4$ cannot support a bound state if simultaneoulsy
\begin{eqnarray}
\label{g-crit-H4}
g_{12}&\le&1-\alpha^2-(m/M)\alpha^2,\nonumber\\
g_{34}&\le&-(M/m)(1-\alpha)^2+\alpha(2-\alpha),\\
g_{mM}&\le&1/2.\nonumber
\end{eqnarray}
Interestingly, the condition on  $g_{mM}$ decouples. Hence the domain
for 4-body binding without 2-body binding consists at most of
$1/2<g_{mM}<1$, and, in the $(g_{12},g_{34})$ plane, the area
between the unit square and a parabola, as shown in Fig.\ \ref{Fig2}.
One should of course exclude the domain corresponding to 3-body
binding to get
a genuine halo.

Our interest in 3-body systems was clearly triggered by nuclei like
$^6$He
or $^{11}$Li with two neutrons weakly attached to a compact nucleus
\cite{Bang}. We assume
a spin singlet state for the two neutrons, so that their spatial wave
function
is symmetric.
States with more than two neutrons in the halo unfortunately
escape direct application of our results, because of the Pauli
principle.
Our result on $(m,m,M,M)$ configurations is perhaps relevant for
$(n,n,\Lambda,\Lambda)$ hypernuclei with strangeness $S=-2$, a field
of intense
theoretical studies \cite{Gal}.

We thank A.S.\ Jensen for calling our attention on halo nuclei
during a Workshop at ECT*, whose hospitality is gratefully
acknowledged,
and for many useful informations.
We also benefitted from discussions with D.\ Brink, A.\ Gal, C.\
Gignoux, W.\
Roberts, B.\ Silvestre-Brac,
and T.T.\ Wu. The Institut de Physique Nucl\'eaire is supported by
Universit\'e Claude Bernard, the Institut des Sciences Nucl\'eaires
by
Universit\'e Joseph Fourier, and both by CNRS--IN2P3.



\begin{figure}
\vglue .5cm
\begin{center}
\setlength{\unitlength}{0.240900pt}
\ifx\plotpoint\undefined\newsavebox{\plotpoint}\fi
\sbox{\plotpoint}{\rule[-0.200pt]{0.400pt}{0.400pt}}%
\begin{picture}(900,731)(0,0)
\font\gnuplot=cmr10 at 10pt
\gnuplot
\sbox{\plotpoint}{\rule[-0.200pt]{0.400pt}{0.400pt}}%
\put(220.0,113.0){\rule[-0.200pt]{148.394pt}{0.400pt}}
\put(220.0,113.0){\rule[-0.200pt]{0.400pt}{143.335pt}}
\put(220.0,113.0){\rule[-0.200pt]{4.818pt}{0.400pt}}
\put(198,113){\makebox(0,0)[r]{0}}
\put(816.0,113.0){\rule[-0.200pt]{4.818pt}{0.400pt}}
\put(220.0,232.0){\rule[-0.200pt]{4.818pt}{0.400pt}}
\put(198,232){\makebox(0,0)[r]{0.2}}
\put(816.0,232.0){\rule[-0.200pt]{4.818pt}{0.400pt}}
\put(220.0,351.0){\rule[-0.200pt]{4.818pt}{0.400pt}}
\put(198,351){\makebox(0,0)[r]{0.4}}
\put(816.0,351.0){\rule[-0.200pt]{4.818pt}{0.400pt}}
\put(220.0,470.0){\rule[-0.200pt]{4.818pt}{0.400pt}}
\put(198,470){\makebox(0,0)[r]{0.6}}
\put(816.0,470.0){\rule[-0.200pt]{4.818pt}{0.400pt}}
\put(220.0,589.0){\rule[-0.200pt]{4.818pt}{0.400pt}}
\put(198,589){\makebox(0,0)[r]{0.8}}
\put(816.0,589.0){\rule[-0.200pt]{4.818pt}{0.400pt}}
\put(220.0,708.0){\rule[-0.200pt]{4.818pt}{0.400pt}}
\put(198,708){\makebox(0,0)[r]{1}}
\put(816.0,708.0){\rule[-0.200pt]{4.818pt}{0.400pt}}
\put(220.0,113.0){\rule[-0.200pt]{0.400pt}{4.818pt}}
\put(220,68){\makebox(0,0){0}}
\put(220.0,688.0){\rule[-0.200pt]{0.400pt}{4.818pt}}
\put(343.0,113.0){\rule[-0.200pt]{0.400pt}{4.818pt}}
\put(343,68){\makebox(0,0){0.2}}
\put(343.0,688.0){\rule[-0.200pt]{0.400pt}{4.818pt}}
\put(466.0,113.0){\rule[-0.200pt]{0.400pt}{4.818pt}}
\put(466,68){\makebox(0,0){0.4}}
\put(466.0,688.0){\rule[-0.200pt]{0.400pt}{4.818pt}}
\put(590.0,113.0){\rule[-0.200pt]{0.400pt}{4.818pt}}
\put(590,68){\makebox(0,0){0.6}}
\put(590.0,688.0){\rule[-0.200pt]{0.400pt}{4.818pt}}
\put(713.0,113.0){\rule[-0.200pt]{0.400pt}{4.818pt}}
\put(713,68){\makebox(0,0){0.8}}
\put(713.0,688.0){\rule[-0.200pt]{0.400pt}{4.818pt}}
\put(836.0,113.0){\rule[-0.200pt]{0.400pt}{4.818pt}}
\put(836,68){\makebox(0,0){1}}
\put(836.0,688.0){\rule[-0.200pt]{0.400pt}{4.818pt}}
\put(220.0,113.0){\rule[-0.200pt]{148.394pt}{0.400pt}}
\put(836.0,113.0){\rule[-0.200pt]{0.400pt}{143.335pt}}
\put(220.0,708.0){\rule[-0.200pt]{148.394pt}{0.400pt}}
\put(67,410){\makebox(0,0){$g$}}
\put(528,-22){\makebox(0,0){$h$}}
\put(861,125){\makebox(0,0)[l]{{A}}}
\put(232,738){\makebox(0,0)[l]{{B}}}
\put(220.0,113.0){\rule[-0.200pt]{0.400pt}{143.335pt}}
\put(836,113){\usebox{\plotpoint}}
\multiput(834.92,113.00)(-0.492,1.013){21}{\rule{0.119pt}{0.900pt}}
\multiput(835.17,113.00)(-12.000,22.132){2}{\rule{0.400pt}{0.450pt}}
\multiput(822.92,137.00)(-0.492,0.927){21}{\rule{0.119pt}{0.833pt}}
\multiput(823.17,137.00)(-12.000,20.270){2}{\rule{0.400pt}{0.417pt}}
\multiput(810.92,159.00)(-0.492,0.927){21}{\rule{0.119pt}{0.833pt}}
\multiput(811.17,159.00)(-12.000,20.270){2}{\rule{0.400pt}{0.417pt}}
\multiput(798.92,181.00)(-0.492,0.967){19}{\rule{0.118pt}{0.864pt}}
\multiput(799.17,181.00)(-11.000,19.207){2}{\rule{0.400pt}{0.432pt}}
\multiput(787.92,202.00)(-0.492,0.920){19}{\rule{0.118pt}{0.827pt}}
\multiput(788.17,202.00)(-11.000,18.283){2}{\rule{0.400pt}{0.414pt}}
\multiput(776.92,222.00)(-0.491,0.964){17}{\rule{0.118pt}{0.860pt}}
\multiput(777.17,222.00)(-10.000,17.215){2}{\rule{0.400pt}{0.430pt}}
\multiput(766.92,241.00)(-0.492,0.873){19}{\rule{0.118pt}{0.791pt}}
\multiput(767.17,241.00)(-11.000,17.358){2}{\rule{0.400pt}{0.395pt}}
\multiput(755.92,260.00)(-0.491,0.912){17}{\rule{0.118pt}{0.820pt}}
\multiput(756.17,260.00)(-10.000,16.298){2}{\rule{0.400pt}{0.410pt}}
\multiput(745.93,278.00)(-0.489,0.961){15}{\rule{0.118pt}{0.856pt}}
\multiput(746.17,278.00)(-9.000,15.224){2}{\rule{0.400pt}{0.428pt}}
\multiput(736.92,295.00)(-0.491,0.808){17}{\rule{0.118pt}{0.740pt}}
\multiput(737.17,295.00)(-10.000,14.464){2}{\rule{0.400pt}{0.370pt}}
\multiput(726.93,311.00)(-0.489,0.902){15}{\rule{0.118pt}{0.811pt}}
\multiput(727.17,311.00)(-9.000,14.316){2}{\rule{0.400pt}{0.406pt}}
\multiput(717.93,327.00)(-0.489,0.844){15}{\rule{0.118pt}{0.767pt}}
\multiput(718.17,327.00)(-9.000,13.409){2}{\rule{0.400pt}{0.383pt}}
\multiput(708.93,342.00)(-0.489,0.844){15}{\rule{0.118pt}{0.767pt}}
\multiput(709.17,342.00)(-9.000,13.409){2}{\rule{0.400pt}{0.383pt}}
\multiput(699.93,357.00)(-0.488,0.890){13}{\rule{0.117pt}{0.800pt}}
\multiput(700.17,357.00)(-8.000,12.340){2}{\rule{0.400pt}{0.400pt}}
\multiput(691.93,371.00)(-0.488,0.890){13}{\rule{0.117pt}{0.800pt}}
\multiput(692.17,371.00)(-8.000,12.340){2}{\rule{0.400pt}{0.400pt}}
\multiput(683.93,385.00)(-0.488,0.824){13}{\rule{0.117pt}{0.750pt}}
\multiput(684.17,385.00)(-8.000,11.443){2}{\rule{0.400pt}{0.375pt}}
\multiput(675.93,398.00)(-0.488,0.824){13}{\rule{0.117pt}{0.750pt}}
\multiput(676.17,398.00)(-8.000,11.443){2}{\rule{0.400pt}{0.375pt}}
\multiput(667.93,411.00)(-0.488,0.758){13}{\rule{0.117pt}{0.700pt}}
\multiput(668.17,411.00)(-8.000,10.547){2}{\rule{0.400pt}{0.350pt}}
\multiput(659.93,423.00)(-0.485,0.874){11}{\rule{0.117pt}{0.786pt}}
\multiput(660.17,423.00)(-7.000,10.369){2}{\rule{0.400pt}{0.393pt}}
\multiput(652.93,435.00)(-0.488,0.692){13}{\rule{0.117pt}{0.650pt}}
\multiput(653.17,435.00)(-8.000,9.651){2}{\rule{0.400pt}{0.325pt}}
\multiput(644.93,446.00)(-0.485,0.798){11}{\rule{0.117pt}{0.729pt}}
\multiput(645.17,446.00)(-7.000,9.488){2}{\rule{0.400pt}{0.364pt}}
\multiput(637.93,457.00)(-0.485,0.721){11}{\rule{0.117pt}{0.671pt}}
\multiput(638.17,457.00)(-7.000,8.606){2}{\rule{0.400pt}{0.336pt}}
\multiput(630.93,467.00)(-0.482,0.852){9}{\rule{0.116pt}{0.767pt}}
\multiput(631.17,467.00)(-6.000,8.409){2}{\rule{0.400pt}{0.383pt}}
\multiput(624.93,477.00)(-0.485,0.721){11}{\rule{0.117pt}{0.671pt}}
\multiput(625.17,477.00)(-7.000,8.606){2}{\rule{0.400pt}{0.336pt}}
\multiput(617.93,487.00)(-0.482,0.762){9}{\rule{0.116pt}{0.700pt}}
\multiput(618.17,487.00)(-6.000,7.547){2}{\rule{0.400pt}{0.350pt}}
\multiput(611.93,496.00)(-0.485,0.645){11}{\rule{0.117pt}{0.614pt}}
\multiput(612.17,496.00)(-7.000,7.725){2}{\rule{0.400pt}{0.307pt}}
\multiput(604.93,505.00)(-0.482,0.762){9}{\rule{0.116pt}{0.700pt}}
\multiput(605.17,505.00)(-6.000,7.547){2}{\rule{0.400pt}{0.350pt}}
\multiput(598.93,514.00)(-0.482,0.671){9}{\rule{0.116pt}{0.633pt}}
\multiput(599.17,514.00)(-6.000,6.685){2}{\rule{0.400pt}{0.317pt}}
\multiput(592.93,522.00)(-0.482,0.671){9}{\rule{0.116pt}{0.633pt}}
\multiput(593.17,522.00)(-6.000,6.685){2}{\rule{0.400pt}{0.317pt}}
\multiput(586.93,530.00)(-0.477,0.821){7}{\rule{0.115pt}{0.740pt}}
\multiput(587.17,530.00)(-5.000,6.464){2}{\rule{0.400pt}{0.370pt}}
\multiput(581.93,538.00)(-0.482,0.581){9}{\rule{0.116pt}{0.567pt}}
\multiput(582.17,538.00)(-6.000,5.824){2}{\rule{0.400pt}{0.283pt}}
\multiput(575.93,545.00)(-0.477,0.710){7}{\rule{0.115pt}{0.660pt}}
\multiput(576.17,545.00)(-5.000,5.630){2}{\rule{0.400pt}{0.330pt}}
\multiput(570.93,552.00)(-0.477,0.710){7}{\rule{0.115pt}{0.660pt}}
\multiput(571.17,552.00)(-5.000,5.630){2}{\rule{0.400pt}{0.330pt}}
\multiput(565.93,559.00)(-0.482,0.581){9}{\rule{0.116pt}{0.567pt}}
\multiput(566.17,559.00)(-6.000,5.824){2}{\rule{0.400pt}{0.283pt}}
\multiput(559.93,566.00)(-0.477,0.599){7}{\rule{0.115pt}{0.580pt}}
\multiput(560.17,566.00)(-5.000,4.796){2}{\rule{0.400pt}{0.290pt}}
\multiput(554.93,572.00)(-0.477,0.599){7}{\rule{0.115pt}{0.580pt}}
\multiput(555.17,572.00)(-5.000,4.796){2}{\rule{0.400pt}{0.290pt}}
\multiput(549.93,578.00)(-0.477,0.599){7}{\rule{0.115pt}{0.580pt}}
\multiput(550.17,578.00)(-5.000,4.796){2}{\rule{0.400pt}{0.290pt}}
\multiput(544.94,584.00)(-0.468,0.774){5}{\rule{0.113pt}{0.700pt}}
\multiput(545.17,584.00)(-4.000,4.547){2}{\rule{0.400pt}{0.350pt}}
\multiput(540.93,590.00)(-0.477,0.599){7}{\rule{0.115pt}{0.580pt}}
\multiput(541.17,590.00)(-5.000,4.796){2}{\rule{0.400pt}{0.290pt}}
\multiput(534.92,596.59)(-0.487,0.477){7}{\rule{0.500pt}{0.115pt}}
\multiput(535.96,595.17)(-3.962,5.000){2}{\rule{0.250pt}{0.400pt}}
\multiput(530.94,601.00)(-0.468,0.627){5}{\rule{0.113pt}{0.600pt}}
\multiput(531.17,601.00)(-4.000,3.755){2}{\rule{0.400pt}{0.300pt}}
\multiput(526.94,606.00)(-0.468,0.627){5}{\rule{0.113pt}{0.600pt}}
\multiput(527.17,606.00)(-4.000,3.755){2}{\rule{0.400pt}{0.300pt}}
\multiput(521.51,611.60)(-0.627,0.468){5}{\rule{0.600pt}{0.113pt}}
\multiput(522.75,610.17)(-3.755,4.000){2}{\rule{0.300pt}{0.400pt}}
\multiput(517.94,615.00)(-0.468,0.627){5}{\rule{0.113pt}{0.600pt}}
\multiput(518.17,615.00)(-4.000,3.755){2}{\rule{0.400pt}{0.300pt}}
\multiput(512.92,620.60)(-0.481,0.468){5}{\rule{0.500pt}{0.113pt}}
\multiput(513.96,619.17)(-2.962,4.000){2}{\rule{0.250pt}{0.400pt}}
\multiput(509.94,624.00)(-0.468,0.627){5}{\rule{0.113pt}{0.600pt}}
\multiput(510.17,624.00)(-4.000,3.755){2}{\rule{0.400pt}{0.300pt}}
\multiput(504.92,629.60)(-0.481,0.468){5}{\rule{0.500pt}{0.113pt}}
\multiput(505.96,628.17)(-2.962,4.000){2}{\rule{0.250pt}{0.400pt}}
\multiput(500.37,633.61)(-0.685,0.447){3}{\rule{0.633pt}{0.108pt}}
\multiput(501.69,632.17)(-2.685,3.000){2}{\rule{0.317pt}{0.400pt}}
\multiput(497.95,636.00)(-0.447,0.685){3}{\rule{0.108pt}{0.633pt}}
\multiput(498.17,636.00)(-3.000,2.685){2}{\rule{0.400pt}{0.317pt}}
\multiput(493.92,640.60)(-0.481,0.468){5}{\rule{0.500pt}{0.113pt}}
\multiput(494.96,639.17)(-2.962,4.000){2}{\rule{0.250pt}{0.400pt}}
\multiput(489.37,644.61)(-0.685,0.447){3}{\rule{0.633pt}{0.108pt}}
\multiput(490.69,643.17)(-2.685,3.000){2}{\rule{0.317pt}{0.400pt}}
\multiput(485.92,647.61)(-0.462,0.447){3}{\rule{0.500pt}{0.108pt}}
\multiput(486.96,646.17)(-1.962,3.000){2}{\rule{0.250pt}{0.400pt}}
\multiput(482.92,650.60)(-0.481,0.468){5}{\rule{0.500pt}{0.113pt}}
\multiput(483.96,649.17)(-2.962,4.000){2}{\rule{0.250pt}{0.400pt}}
\multiput(478.92,654.61)(-0.462,0.447){3}{\rule{0.500pt}{0.108pt}}
\multiput(479.96,653.17)(-1.962,3.000){2}{\rule{0.250pt}{0.400pt}}
\put(475,657.17){\rule{0.700pt}{0.400pt}}
\multiput(476.55,656.17)(-1.547,2.000){2}{\rule{0.350pt}{0.400pt}}
\multiput(472.37,659.61)(-0.685,0.447){3}{\rule{0.633pt}{0.108pt}}
\multiput(473.69,658.17)(-2.685,3.000){2}{\rule{0.317pt}{0.400pt}}
\multiput(468.92,662.61)(-0.462,0.447){3}{\rule{0.500pt}{0.108pt}}
\multiput(469.96,661.17)(-1.962,3.000){2}{\rule{0.250pt}{0.400pt}}
\put(465,665.17){\rule{0.700pt}{0.400pt}}
\multiput(466.55,664.17)(-1.547,2.000){2}{\rule{0.350pt}{0.400pt}}
\multiput(462.92,667.61)(-0.462,0.447){3}{\rule{0.500pt}{0.108pt}}
\multiput(463.96,666.17)(-1.962,3.000){2}{\rule{0.250pt}{0.400pt}}
\put(459,670.17){\rule{0.700pt}{0.400pt}}
\multiput(460.55,669.17)(-1.547,2.000){2}{\rule{0.350pt}{0.400pt}}
\put(456,672.17){\rule{0.700pt}{0.400pt}}
\multiput(457.55,671.17)(-1.547,2.000){2}{\rule{0.350pt}{0.400pt}}
\multiput(453.92,674.61)(-0.462,0.447){3}{\rule{0.500pt}{0.108pt}}
\multiput(454.96,673.17)(-1.962,3.000){2}{\rule{0.250pt}{0.400pt}}
\put(450,677.17){\rule{0.700pt}{0.400pt}}
\multiput(451.55,676.17)(-1.547,2.000){2}{\rule{0.350pt}{0.400pt}}
\put(447,679.17){\rule{0.700pt}{0.400pt}}
\multiput(448.55,678.17)(-1.547,2.000){2}{\rule{0.350pt}{0.400pt}}
\put(444,680.67){\rule{0.723pt}{0.400pt}}
\multiput(445.50,680.17)(-1.500,1.000){2}{\rule{0.361pt}{0.400pt}}
\put(442,682.17){\rule{0.482pt}{0.400pt}}
\multiput(443.00,681.17)(-1.000,2.000){2}{\rule{0.241pt}{0.400pt}}
\put(439,684.17){\rule{0.700pt}{0.400pt}}
\multiput(440.55,683.17)(-1.547,2.000){2}{\rule{0.350pt}{0.400pt}}
\put(436,685.67){\rule{0.723pt}{0.400pt}}
\multiput(437.50,685.17)(-1.500,1.000){2}{\rule{0.361pt}{0.400pt}}
\put(434,687.17){\rule{0.482pt}{0.400pt}}
\multiput(435.00,686.17)(-1.000,2.000){2}{\rule{0.241pt}{0.400pt}}
\put(431,688.67){\rule{0.723pt}{0.400pt}}
\multiput(432.50,688.17)(-1.500,1.000){2}{\rule{0.361pt}{0.400pt}}
\put(429,690.17){\rule{0.482pt}{0.400pt}}
\multiput(430.00,689.17)(-1.000,2.000){2}{\rule{0.241pt}{0.400pt}}
\put(426,691.67){\rule{0.723pt}{0.400pt}}
\multiput(427.50,691.17)(-1.500,1.000){2}{\rule{0.361pt}{0.400pt}}
\put(424,692.67){\rule{0.482pt}{0.400pt}}
\multiput(425.00,692.17)(-1.000,1.000){2}{\rule{0.241pt}{0.400pt}}
\put(422,694.17){\rule{0.482pt}{0.400pt}}
\multiput(423.00,693.17)(-1.000,2.000){2}{\rule{0.241pt}{0.400pt}}
\put(419,695.67){\rule{0.723pt}{0.400pt}}
\multiput(420.50,695.17)(-1.500,1.000){2}{\rule{0.361pt}{0.400pt}}
\put(417,696.67){\rule{0.482pt}{0.400pt}}
\multiput(418.00,696.17)(-1.000,1.000){2}{\rule{0.241pt}{0.400pt}}
\put(415,697.67){\rule{0.482pt}{0.400pt}}
\multiput(416.00,697.17)(-1.000,1.000){2}{\rule{0.241pt}{0.400pt}}
\put(413,698.67){\rule{0.482pt}{0.400pt}}
\multiput(414.00,698.17)(-1.000,1.000){2}{\rule{0.241pt}{0.400pt}}
\put(408,699.67){\rule{0.723pt}{0.400pt}}
\multiput(409.50,699.17)(-1.500,1.000){2}{\rule{0.361pt}{0.400pt}}
\put(406,700.67){\rule{0.482pt}{0.400pt}}
\multiput(407.00,700.17)(-1.000,1.000){2}{\rule{0.241pt}{0.400pt}}
\put(404,701.67){\rule{0.482pt}{0.400pt}}
\multiput(405.00,701.17)(-1.000,1.000){2}{\rule{0.241pt}{0.400pt}}
\put(411.0,700.0){\rule[-0.200pt]{0.482pt}{0.400pt}}
\put(400,702.67){\rule{0.482pt}{0.400pt}}
\multiput(401.00,702.17)(-1.000,1.000){2}{\rule{0.241pt}{0.400pt}}
\put(398,703.67){\rule{0.482pt}{0.400pt}}
\multiput(399.00,703.17)(-1.000,1.000){2}{\rule{0.241pt}{0.400pt}}
\put(402.0,703.0){\rule[-0.200pt]{0.482pt}{0.400pt}}
\put(395,704.67){\rule{0.241pt}{0.400pt}}
\multiput(395.50,704.17)(-0.500,1.000){2}{\rule{0.120pt}{0.400pt}}
\put(396.0,705.0){\rule[-0.200pt]{0.482pt}{0.400pt}}
\put(389,705.67){\rule{0.482pt}{0.400pt}}
\multiput(390.00,705.17)(-1.000,1.000){2}{\rule{0.241pt}{0.400pt}}
\put(391.0,706.0){\rule[-0.200pt]{0.964pt}{0.400pt}}
\put(382,706.67){\rule{0.482pt}{0.400pt}}
\multiput(383.00,706.17)(-1.000,1.000){2}{\rule{0.241pt}{0.400pt}}
\put(384.0,707.0){\rule[-0.200pt]{1.204pt}{0.400pt}}
\put(374.0,708.0){\rule[-0.200pt]{1.927pt}{0.400pt}}
\put(220,708){\usebox{\plotpoint}}
\multiput(220.00,706.93)(0.491,-0.482){9}{\rule{0.500pt}{0.116pt}}
\multiput(220.00,707.17)(4.962,-6.000){2}{\rule{0.250pt}{0.400pt}}
\multiput(226.00,700.93)(0.491,-0.482){9}{\rule{0.500pt}{0.116pt}}
\multiput(226.00,701.17)(4.962,-6.000){2}{\rule{0.250pt}{0.400pt}}
\multiput(232.00,694.93)(0.581,-0.482){9}{\rule{0.567pt}{0.116pt}}
\multiput(232.00,695.17)(5.824,-6.000){2}{\rule{0.283pt}{0.400pt}}
\multiput(239.00,688.93)(0.491,-0.482){9}{\rule{0.500pt}{0.116pt}}
\multiput(239.00,689.17)(4.962,-6.000){2}{\rule{0.250pt}{0.400pt}}
\multiput(245.00,682.93)(0.491,-0.482){9}{\rule{0.500pt}{0.116pt}}
\multiput(245.00,683.17)(4.962,-6.000){2}{\rule{0.250pt}{0.400pt}}
\multiput(251.00,676.93)(0.491,-0.482){9}{\rule{0.500pt}{0.116pt}}
\multiput(251.00,677.17)(4.962,-6.000){2}{\rule{0.250pt}{0.400pt}}
\multiput(257.00,670.93)(0.581,-0.482){9}{\rule{0.567pt}{0.116pt}}
\multiput(257.00,671.17)(5.824,-6.000){2}{\rule{0.283pt}{0.400pt}}
\multiput(264.00,664.93)(0.491,-0.482){9}{\rule{0.500pt}{0.116pt}}
\multiput(264.00,665.17)(4.962,-6.000){2}{\rule{0.250pt}{0.400pt}}
\multiput(270.00,658.93)(0.491,-0.482){9}{\rule{0.500pt}{0.116pt}}
\multiput(270.00,659.17)(4.962,-6.000){2}{\rule{0.250pt}{0.400pt}}
\multiput(276.00,652.93)(0.491,-0.482){9}{\rule{0.500pt}{0.116pt}}
\multiput(276.00,653.17)(4.962,-6.000){2}{\rule{0.250pt}{0.400pt}}
\multiput(282.00,646.93)(0.491,-0.482){9}{\rule{0.500pt}{0.116pt}}
\multiput(282.00,647.17)(4.962,-6.000){2}{\rule{0.250pt}{0.400pt}}
\multiput(288.00,640.93)(0.581,-0.482){9}{\rule{0.567pt}{0.116pt}}
\multiput(288.00,641.17)(5.824,-6.000){2}{\rule{0.283pt}{0.400pt}}
\multiput(295.00,634.93)(0.491,-0.482){9}{\rule{0.500pt}{0.116pt}}
\multiput(295.00,635.17)(4.962,-6.000){2}{\rule{0.250pt}{0.400pt}}
\multiput(301.00,628.93)(0.491,-0.482){9}{\rule{0.500pt}{0.116pt}}
\multiput(301.00,629.17)(4.962,-6.000){2}{\rule{0.250pt}{0.400pt}}
\multiput(307.00,622.93)(0.491,-0.482){9}{\rule{0.500pt}{0.116pt}}
\multiput(307.00,623.17)(4.962,-6.000){2}{\rule{0.250pt}{0.400pt}}
\multiput(313.00,616.93)(0.581,-0.482){9}{\rule{0.567pt}{0.116pt}}
\multiput(313.00,617.17)(5.824,-6.000){2}{\rule{0.283pt}{0.400pt}}
\multiput(320.00,610.93)(0.491,-0.482){9}{\rule{0.500pt}{0.116pt}}
\multiput(320.00,611.17)(4.962,-6.000){2}{\rule{0.250pt}{0.400pt}}
\multiput(326.00,604.93)(0.491,-0.482){9}{\rule{0.500pt}{0.116pt}}
\multiput(326.00,605.17)(4.962,-6.000){2}{\rule{0.250pt}{0.400pt}}
\multiput(332.00,598.93)(0.491,-0.482){9}{\rule{0.500pt}{0.116pt}}
\multiput(332.00,599.17)(4.962,-6.000){2}{\rule{0.250pt}{0.400pt}}
\multiput(338.00,592.93)(0.491,-0.482){9}{\rule{0.500pt}{0.116pt}}
\multiput(338.00,593.17)(4.962,-6.000){2}{\rule{0.250pt}{0.400pt}}
\multiput(344.00,586.93)(0.581,-0.482){9}{\rule{0.567pt}{0.116pt}}
\multiput(344.00,587.17)(5.824,-6.000){2}{\rule{0.283pt}{0.400pt}}
\multiput(351.00,580.93)(0.491,-0.482){9}{\rule{0.500pt}{0.116pt}}
\multiput(351.00,581.17)(4.962,-6.000){2}{\rule{0.250pt}{0.400pt}}
\multiput(357.00,574.93)(0.491,-0.482){9}{\rule{0.500pt}{0.116pt}}
\multiput(357.00,575.17)(4.962,-6.000){2}{\rule{0.250pt}{0.400pt}}
\multiput(363.00,568.93)(0.491,-0.482){9}{\rule{0.500pt}{0.116pt}}
\multiput(363.00,569.17)(4.962,-6.000){2}{\rule{0.250pt}{0.400pt}}
\multiput(369.00,562.93)(0.581,-0.482){9}{\rule{0.567pt}{0.116pt}}
\multiput(369.00,563.17)(5.824,-6.000){2}{\rule{0.283pt}{0.400pt}}
\multiput(376.00,556.93)(0.491,-0.482){9}{\rule{0.500pt}{0.116pt}}
\multiput(376.00,557.17)(4.962,-6.000){2}{\rule{0.250pt}{0.400pt}}
\multiput(382.00,550.93)(0.491,-0.482){9}{\rule{0.500pt}{0.116pt}}
\multiput(382.00,551.17)(4.962,-6.000){2}{\rule{0.250pt}{0.400pt}}
\multiput(388.00,544.93)(0.491,-0.482){9}{\rule{0.500pt}{0.116pt}}
\multiput(388.00,545.17)(4.962,-6.000){2}{\rule{0.250pt}{0.400pt}}
\multiput(394.00,538.93)(0.491,-0.482){9}{\rule{0.500pt}{0.116pt}}
\multiput(394.00,539.17)(4.962,-6.000){2}{\rule{0.250pt}{0.400pt}}
\multiput(400.00,532.93)(0.581,-0.482){9}{\rule{0.567pt}{0.116pt}}
\multiput(400.00,533.17)(5.824,-6.000){2}{\rule{0.283pt}{0.400pt}}
\multiput(407.00,526.93)(0.491,-0.482){9}{\rule{0.500pt}{0.116pt}}
\multiput(407.00,527.17)(4.962,-6.000){2}{\rule{0.250pt}{0.400pt}}
\multiput(413.00,520.93)(0.491,-0.482){9}{\rule{0.500pt}{0.116pt}}
\multiput(413.00,521.17)(4.962,-6.000){2}{\rule{0.250pt}{0.400pt}}
\multiput(419.00,514.93)(0.491,-0.482){9}{\rule{0.500pt}{0.116pt}}
\multiput(419.00,515.17)(4.962,-6.000){2}{\rule{0.250pt}{0.400pt}}
\multiput(425.00,508.93)(0.581,-0.482){9}{\rule{0.567pt}{0.116pt}}
\multiput(425.00,509.17)(5.824,-6.000){2}{\rule{0.283pt}{0.400pt}}
\multiput(432.00,502.93)(0.491,-0.482){9}{\rule{0.500pt}{0.116pt}}
\multiput(432.00,503.17)(4.962,-6.000){2}{\rule{0.250pt}{0.400pt}}
\multiput(438.00,496.93)(0.491,-0.482){9}{\rule{0.500pt}{0.116pt}}
\multiput(438.00,497.17)(4.962,-6.000){2}{\rule{0.250pt}{0.400pt}}
\multiput(444.00,490.93)(0.491,-0.482){9}{\rule{0.500pt}{0.116pt}}
\multiput(444.00,491.17)(4.962,-6.000){2}{\rule{0.250pt}{0.400pt}}
\multiput(450.00,484.93)(0.491,-0.482){9}{\rule{0.500pt}{0.116pt}}
\multiput(450.00,485.17)(4.962,-6.000){2}{\rule{0.250pt}{0.400pt}}
\multiput(456.00,478.93)(0.581,-0.482){9}{\rule{0.567pt}{0.116pt}}
\multiput(456.00,479.17)(5.824,-6.000){2}{\rule{0.283pt}{0.400pt}}
\multiput(463.00,472.93)(0.491,-0.482){9}{\rule{0.500pt}{0.116pt}}
\multiput(463.00,473.17)(4.962,-6.000){2}{\rule{0.250pt}{0.400pt}}
\multiput(469.00,466.93)(0.491,-0.482){9}{\rule{0.500pt}{0.116pt}}
\multiput(469.00,467.17)(4.962,-6.000){2}{\rule{0.250pt}{0.400pt}}
\multiput(475.00,460.93)(0.491,-0.482){9}{\rule{0.500pt}{0.116pt}}
\multiput(475.00,461.17)(4.962,-6.000){2}{\rule{0.250pt}{0.400pt}}
\multiput(481.00,454.93)(0.581,-0.482){9}{\rule{0.567pt}{0.116pt}}
\multiput(481.00,455.17)(5.824,-6.000){2}{\rule{0.283pt}{0.400pt}}
\multiput(488.00,448.93)(0.491,-0.482){9}{\rule{0.500pt}{0.116pt}}
\multiput(488.00,449.17)(4.962,-6.000){2}{\rule{0.250pt}{0.400pt}}
\multiput(494.00,442.93)(0.491,-0.482){9}{\rule{0.500pt}{0.116pt}}
\multiput(494.00,443.17)(4.962,-6.000){2}{\rule{0.250pt}{0.400pt}}
\multiput(500.00,436.93)(0.491,-0.482){9}{\rule{0.500pt}{0.116pt}}
\multiput(500.00,437.17)(4.962,-6.000){2}{\rule{0.250pt}{0.400pt}}
\multiput(506.00,430.93)(0.491,-0.482){9}{\rule{0.500pt}{0.116pt}}
\multiput(506.00,431.17)(4.962,-6.000){2}{\rule{0.250pt}{0.400pt}}
\multiput(512.00,424.93)(0.581,-0.482){9}{\rule{0.567pt}{0.116pt}}
\multiput(512.00,425.17)(5.824,-6.000){2}{\rule{0.283pt}{0.400pt}}
\multiput(519.00,418.93)(0.491,-0.482){9}{\rule{0.500pt}{0.116pt}}
\multiput(519.00,419.17)(4.962,-6.000){2}{\rule{0.250pt}{0.400pt}}
\multiput(525.59,411.65)(0.482,-0.581){9}{\rule{0.116pt}{0.567pt}}
\multiput(524.17,412.82)(6.000,-5.824){2}{\rule{0.400pt}{0.283pt}}
\multiput(531.00,405.93)(0.491,-0.482){9}{\rule{0.500pt}{0.116pt}}
\multiput(531.00,406.17)(4.962,-6.000){2}{\rule{0.250pt}{0.400pt}}
\multiput(537.00,399.93)(0.581,-0.482){9}{\rule{0.567pt}{0.116pt}}
\multiput(537.00,400.17)(5.824,-6.000){2}{\rule{0.283pt}{0.400pt}}
\multiput(544.00,393.93)(0.491,-0.482){9}{\rule{0.500pt}{0.116pt}}
\multiput(544.00,394.17)(4.962,-6.000){2}{\rule{0.250pt}{0.400pt}}
\multiput(550.00,387.93)(0.491,-0.482){9}{\rule{0.500pt}{0.116pt}}
\multiput(550.00,388.17)(4.962,-6.000){2}{\rule{0.250pt}{0.400pt}}
\multiput(556.00,381.93)(0.491,-0.482){9}{\rule{0.500pt}{0.116pt}}
\multiput(556.00,382.17)(4.962,-6.000){2}{\rule{0.250pt}{0.400pt}}
\multiput(562.00,375.93)(0.491,-0.482){9}{\rule{0.500pt}{0.116pt}}
\multiput(562.00,376.17)(4.962,-6.000){2}{\rule{0.250pt}{0.400pt}}
\multiput(568.00,369.93)(0.581,-0.482){9}{\rule{0.567pt}{0.116pt}}
\multiput(568.00,370.17)(5.824,-6.000){2}{\rule{0.283pt}{0.400pt}}
\multiput(575.00,363.93)(0.491,-0.482){9}{\rule{0.500pt}{0.116pt}}
\multiput(575.00,364.17)(4.962,-6.000){2}{\rule{0.250pt}{0.400pt}}
\multiput(581.00,357.93)(0.491,-0.482){9}{\rule{0.500pt}{0.116pt}}
\multiput(581.00,358.17)(4.962,-6.000){2}{\rule{0.250pt}{0.400pt}}
\multiput(587.00,351.93)(0.491,-0.482){9}{\rule{0.500pt}{0.116pt}}
\multiput(587.00,352.17)(4.962,-6.000){2}{\rule{0.250pt}{0.400pt}}
\multiput(593.00,345.93)(0.581,-0.482){9}{\rule{0.567pt}{0.116pt}}
\multiput(593.00,346.17)(5.824,-6.000){2}{\rule{0.283pt}{0.400pt}}
\multiput(600.00,339.93)(0.491,-0.482){9}{\rule{0.500pt}{0.116pt}}
\multiput(600.00,340.17)(4.962,-6.000){2}{\rule{0.250pt}{0.400pt}}
\multiput(606.00,333.93)(0.491,-0.482){9}{\rule{0.500pt}{0.116pt}}
\multiput(606.00,334.17)(4.962,-6.000){2}{\rule{0.250pt}{0.400pt}}
\multiput(612.00,327.93)(0.491,-0.482){9}{\rule{0.500pt}{0.116pt}}
\multiput(612.00,328.17)(4.962,-6.000){2}{\rule{0.250pt}{0.400pt}}
\multiput(618.00,321.93)(0.491,-0.482){9}{\rule{0.500pt}{0.116pt}}
\multiput(618.00,322.17)(4.962,-6.000){2}{\rule{0.250pt}{0.400pt}}
\multiput(624.00,315.93)(0.581,-0.482){9}{\rule{0.567pt}{0.116pt}}
\multiput(624.00,316.17)(5.824,-6.000){2}{\rule{0.283pt}{0.400pt}}
\multiput(631.00,309.93)(0.491,-0.482){9}{\rule{0.500pt}{0.116pt}}
\multiput(631.00,310.17)(4.962,-6.000){2}{\rule{0.250pt}{0.400pt}}
\multiput(637.00,303.93)(0.491,-0.482){9}{\rule{0.500pt}{0.116pt}}
\multiput(637.00,304.17)(4.962,-6.000){2}{\rule{0.250pt}{0.400pt}}
\multiput(643.00,297.93)(0.491,-0.482){9}{\rule{0.500pt}{0.116pt}}
\multiput(643.00,298.17)(4.962,-6.000){2}{\rule{0.250pt}{0.400pt}}
\multiput(649.00,291.93)(0.581,-0.482){9}{\rule{0.567pt}{0.116pt}}
\multiput(649.00,292.17)(5.824,-6.000){2}{\rule{0.283pt}{0.400pt}}
\multiput(656.00,285.93)(0.491,-0.482){9}{\rule{0.500pt}{0.116pt}}
\multiput(656.00,286.17)(4.962,-6.000){2}{\rule{0.250pt}{0.400pt}}
\multiput(662.00,279.93)(0.491,-0.482){9}{\rule{0.500pt}{0.116pt}}
\multiput(662.00,280.17)(4.962,-6.000){2}{\rule{0.250pt}{0.400pt}}
\multiput(668.00,273.93)(0.491,-0.482){9}{\rule{0.500pt}{0.116pt}}
\multiput(668.00,274.17)(4.962,-6.000){2}{\rule{0.250pt}{0.400pt}}
\multiput(674.00,267.93)(0.491,-0.482){9}{\rule{0.500pt}{0.116pt}}
\multiput(674.00,268.17)(4.962,-6.000){2}{\rule{0.250pt}{0.400pt}}
\multiput(680.00,261.93)(0.581,-0.482){9}{\rule{0.567pt}{0.116pt}}
\multiput(680.00,262.17)(5.824,-6.000){2}{\rule{0.283pt}{0.400pt}}
\multiput(687.00,255.93)(0.491,-0.482){9}{\rule{0.500pt}{0.116pt}}
\multiput(687.00,256.17)(4.962,-6.000){2}{\rule{0.250pt}{0.400pt}}
\multiput(693.00,249.93)(0.491,-0.482){9}{\rule{0.500pt}{0.116pt}}
\multiput(693.00,250.17)(4.962,-6.000){2}{\rule{0.250pt}{0.400pt}}
\multiput(699.00,243.93)(0.491,-0.482){9}{\rule{0.500pt}{0.116pt}}
\multiput(699.00,244.17)(4.962,-6.000){2}{\rule{0.250pt}{0.400pt}}
\multiput(705.00,237.93)(0.581,-0.482){9}{\rule{0.567pt}{0.116pt}}
\multiput(705.00,238.17)(5.824,-6.000){2}{\rule{0.283pt}{0.400pt}}
\multiput(712.00,231.93)(0.491,-0.482){9}{\rule{0.500pt}{0.116pt}}
\multiput(712.00,232.17)(4.962,-6.000){2}{\rule{0.250pt}{0.400pt}}
\multiput(718.00,225.93)(0.491,-0.482){9}{\rule{0.500pt}{0.116pt}}
\multiput(718.00,226.17)(4.962,-6.000){2}{\rule{0.250pt}{0.400pt}}
\multiput(724.00,219.93)(0.491,-0.482){9}{\rule{0.500pt}{0.116pt}}
\multiput(724.00,220.17)(4.962,-6.000){2}{\rule{0.250pt}{0.400pt}}
\multiput(730.00,213.93)(0.491,-0.482){9}{\rule{0.500pt}{0.116pt}}
\multiput(730.00,214.17)(4.962,-6.000){2}{\rule{0.250pt}{0.400pt}}
\multiput(736.00,207.93)(0.581,-0.482){9}{\rule{0.567pt}{0.116pt}}
\multiput(736.00,208.17)(5.824,-6.000){2}{\rule{0.283pt}{0.400pt}}
\multiput(743.00,201.93)(0.491,-0.482){9}{\rule{0.500pt}{0.116pt}}
\multiput(743.00,202.17)(4.962,-6.000){2}{\rule{0.250pt}{0.400pt}}
\multiput(749.00,195.93)(0.491,-0.482){9}{\rule{0.500pt}{0.116pt}}
\multiput(749.00,196.17)(4.962,-6.000){2}{\rule{0.250pt}{0.400pt}}
\multiput(755.00,189.93)(0.491,-0.482){9}{\rule{0.500pt}{0.116pt}}
\multiput(755.00,190.17)(4.962,-6.000){2}{\rule{0.250pt}{0.400pt}}
\multiput(761.00,183.93)(0.581,-0.482){9}{\rule{0.567pt}{0.116pt}}
\multiput(761.00,184.17)(5.824,-6.000){2}{\rule{0.283pt}{0.400pt}}
\multiput(768.00,177.93)(0.491,-0.482){9}{\rule{0.500pt}{0.116pt}}
\multiput(768.00,178.17)(4.962,-6.000){2}{\rule{0.250pt}{0.400pt}}
\multiput(774.00,171.93)(0.491,-0.482){9}{\rule{0.500pt}{0.116pt}}
\multiput(774.00,172.17)(4.962,-6.000){2}{\rule{0.250pt}{0.400pt}}
\multiput(780.00,165.93)(0.491,-0.482){9}{\rule{0.500pt}{0.116pt}}
\multiput(780.00,166.17)(4.962,-6.000){2}{\rule{0.250pt}{0.400pt}}
\multiput(786.00,159.93)(0.491,-0.482){9}{\rule{0.500pt}{0.116pt}}
\multiput(786.00,160.17)(4.962,-6.000){2}{\rule{0.250pt}{0.400pt}}
\multiput(792.00,153.93)(0.581,-0.482){9}{\rule{0.567pt}{0.116pt}}
\multiput(792.00,154.17)(5.824,-6.000){2}{\rule{0.283pt}{0.400pt}}
\multiput(799.00,147.93)(0.491,-0.482){9}{\rule{0.500pt}{0.116pt}}
\multiput(799.00,148.17)(4.962,-6.000){2}{\rule{0.250pt}{0.400pt}}
\multiput(805.00,141.93)(0.491,-0.482){9}{\rule{0.500pt}{0.116pt}}
\multiput(805.00,142.17)(4.962,-6.000){2}{\rule{0.250pt}{0.400pt}}
\multiput(811.00,135.93)(0.491,-0.482){9}{\rule{0.500pt}{0.116pt}}
\multiput(811.00,136.17)(4.962,-6.000){2}{\rule{0.250pt}{0.400pt}}
\multiput(817.00,129.93)(0.581,-0.482){9}{\rule{0.567pt}{0.116pt}}
\multiput(817.00,130.17)(5.824,-6.000){2}{\rule{0.283pt}{0.400pt}}
\multiput(824.00,123.93)(0.491,-0.482){9}{\rule{0.500pt}{0.116pt}}
\multiput(824.00,124.17)(4.962,-6.000){2}{\rule{0.250pt}{0.400pt}}
\multiput(830.00,117.93)(0.491,-0.482){9}{\rule{0.500pt}{0.116pt}}
\multiput(830.00,118.17)(4.962,-6.000){2}{\rule{0.250pt}{0.400pt}}
\sbox{\plotpoint}{\rule[-0.500pt]{1.000pt}{1.000pt}}%
\put(836,113){\usebox{\plotpoint}}
\put(836.00,113.00){\usebox{\plotpoint}}
\put(821.21,127.56){\usebox{\plotpoint}}
\put(805.91,141.58){\usebox{\plotpoint}}
\put(790.57,155.57){\usebox{\plotpoint}}
\multiput(789,157)(-15.358,13.962){0}{\usebox{\plotpoint}}
\put(775.34,169.66){\usebox{\plotpoint}}
\put(759.97,183.57){\usebox{\plotpoint}}
\multiput(757,186)(-15.427,13.885){0}{\usebox{\plotpoint}}
\put(744.55,197.45){\usebox{\plotpoint}}
\put(729.03,211.18){\usebox{\plotpoint}}
\multiput(728,212)(-15.513,13.789){0}{\usebox{\plotpoint}}
\put(713.47,224.92){\usebox{\plotpoint}}
\multiput(710,228)(-16.383,12.743){0}{\usebox{\plotpoint}}
\put(697.45,238.10){\usebox{\plotpoint}}
\multiput(693,242)(-15.620,13.668){0}{\usebox{\plotpoint}}
\put(681.83,251.77){\usebox{\plotpoint}}
\multiput(677,256)(-16.604,12.453){0}{\usebox{\plotpoint}}
\put(665.53,264.60){\usebox{\plotpoint}}
\multiput(661,268)(-15.759,13.508){0}{\usebox{\plotpoint}}
\put(649.02,277.11){\usebox{\plotpoint}}
\multiput(646,279)(-15.759,13.508){0}{\usebox{\plotpoint}}
\put(632.51,289.63){\usebox{\plotpoint}}
\multiput(632,290)(-15.945,13.287){0}{\usebox{\plotpoint}}
\multiput(626,295)(-16.889,12.064){0}{\usebox{\plotpoint}}
\put(616.15,302.38){\usebox{\plotpoint}}
\multiput(613,305)(-18.021,10.298){0}{\usebox{\plotpoint}}
\multiput(606,309)(-17.270,11.513){0}{\usebox{\plotpoint}}
\put(598.94,313.89){\usebox{\plotpoint}}
\multiput(594,318)(-17.270,11.513){0}{\usebox{\plotpoint}}
\multiput(588,322)(-17.798,10.679){0}{\usebox{\plotpoint}}
\put(581.93,325.71){\usebox{\plotpoint}}
\multiput(577,329)(-16.207,12.966){0}{\usebox{\plotpoint}}
\multiput(572,333)(-17.798,10.679){0}{\usebox{\plotpoint}}
\put(564.68,337.16){\usebox{\plotpoint}}
\multiput(561,339)(-16.207,12.966){0}{\usebox{\plotpoint}}
\multiput(556,343)(-17.798,10.679){0}{\usebox{\plotpoint}}
\put(547.22,348.27){\usebox{\plotpoint}}
\multiput(546,349)(-16.604,12.453){0}{\usebox{\plotpoint}}
\multiput(542,352)(-19.271,7.708){0}{\usebox{\plotpoint}}
\multiput(537,354)(-17.798,10.679){0}{\usebox{\plotpoint}}
\put(529.21,358.39){\usebox{\plotpoint}}
\multiput(528,359)(-16.604,12.453){0}{\usebox{\plotpoint}}
\multiput(524,362)(-19.271,7.708){0}{\usebox{\plotpoint}}
\multiput(519,364)(-18.564,9.282){0}{\usebox{\plotpoint}}
\put(511.36,368.73){\usebox{\plotpoint}}
\multiput(511,369)(-18.564,9.282){0}{\usebox{\plotpoint}}
\multiput(507,371)(-18.564,9.282){0}{\usebox{\plotpoint}}
\multiput(503,373)(-18.564,9.282){0}{\usebox{\plotpoint}}
\multiput(499,375)(-17.270,11.513){0}{\usebox{\plotpoint}}
\put(492.82,377.79){\usebox{\plotpoint}}
\multiput(492,378)(-18.564,9.282){0}{\usebox{\plotpoint}}
\multiput(488,380)(-17.270,11.513){0}{\usebox{\plotpoint}}
\multiput(485,382)(-20.136,5.034){0}{\usebox{\plotpoint}}
\multiput(481,383)(-17.270,11.513){0}{\usebox{\plotpoint}}
\multiput(478,385)(-19.690,6.563){0}{\usebox{\plotpoint}}
\put(474.16,386.42){\usebox{\plotpoint}}
\multiput(471,388)(-19.690,6.563){0}{\usebox{\plotpoint}}
\multiput(468,389)(-19.690,6.563){0}{\usebox{\plotpoint}}
\multiput(465,390)(-19.690,6.563){0}{\usebox{\plotpoint}}
\multiput(462,391)(-17.270,11.513){0}{\usebox{\plotpoint}}
\multiput(459,393)(-19.690,6.563){0}{\usebox{\plotpoint}}
\put(455.08,394.31){\usebox{\plotpoint}}
\multiput(453,395)(-19.690,6.563){0}{\usebox{\plotpoint}}
\multiput(450,396)(-19.690,6.563){0}{\usebox{\plotpoint}}
\multiput(447,397)(-19.690,6.563){0}{\usebox{\plotpoint}}
\multiput(444,398)(-18.564,9.282){0}{\usebox{\plotpoint}}
\multiput(442,399)(-20.756,0.000){0}{\usebox{\plotpoint}}
\multiput(439,399)(-19.690,6.563){0}{\usebox{\plotpoint}}
\put(435.39,400.30){\usebox{\plotpoint}}
\multiput(434,401)(-19.690,6.563){0}{\usebox{\plotpoint}}
\multiput(431,402)(-20.756,0.000){0}{\usebox{\plotpoint}}
\multiput(429,402)(-19.690,6.563){0}{\usebox{\plotpoint}}
\multiput(426,403)(-18.564,9.282){0}{\usebox{\plotpoint}}
\multiput(424,404)(-20.756,0.000){0}{\usebox{\plotpoint}}
\multiput(422,404)(-19.690,6.563){0}{\usebox{\plotpoint}}
\multiput(419,405)(-20.756,0.000){0}{\usebox{\plotpoint}}
\put(415.68,405.66){\usebox{\plotpoint}}
\multiput(415,406)(-20.756,0.000){0}{\usebox{\plotpoint}}
\multiput(413,406)(-18.564,9.282){0}{\usebox{\plotpoint}}
\multiput(411,407)(-20.756,0.000){0}{\usebox{\plotpoint}}
\multiput(408,407)(-18.564,9.282){0}{\usebox{\plotpoint}}
\multiput(406,408)(-20.756,0.000){0}{\usebox{\plotpoint}}
\multiput(404,408)(-20.756,0.000){0}{\usebox{\plotpoint}}
\multiput(402,408)(-18.564,9.282){0}{\usebox{\plotpoint}}
\multiput(400,409)(-20.756,0.000){0}{\usebox{\plotpoint}}
\multiput(398,409)(-20.756,0.000){0}{\usebox{\plotpoint}}
\put(395.71,409.00){\usebox{\plotpoint}}
\multiput(395,409)(-20.756,0.000){0}{\usebox{\plotpoint}}
\multiput(393,409)(-18.564,9.282){0}{\usebox{\plotpoint}}
\multiput(391,410)(-20.756,0.000){0}{\usebox{\plotpoint}}
\multiput(389,410)(-20.756,0.000){0}{\usebox{\plotpoint}}
\multiput(387,410)(-20.756,0.000){0}{\usebox{\plotpoint}}
\multiput(385,410)(-20.756,0.000){0}{\usebox{\plotpoint}}
\multiput(384,410)(-20.756,0.000){0}{\usebox{\plotpoint}}
\multiput(382,410)(-20.756,0.000){0}{\usebox{\plotpoint}}
\multiput(380,410)(-20.756,0.000){0}{\usebox{\plotpoint}}
\multiput(379,410)(-20.756,0.000){0}{\usebox{\plotpoint}}
\multiput(377,410)(-20.756,0.000){0}{\usebox{\plotpoint}}
\put(375.28,410.36){\usebox{\plotpoint}}
\put(374,411){\usebox{\plotpoint}}
\end{picture}
\end{center}
\vglue 1cm
\caption{\label{Fig1}Lower limit for the domain of halo binding of
two bosons
in the field of a static source. The coupling to the source is
denoted $h$,
while  $g$ is the interparticle coupling. Two-body binding occurs for
$h>1$ or
$g>1$.
The straight line AB is deduced from the simple decomposition
(\protect{\ref{H2-decomp-1}}), the parabola from
(\protect{\ref{H2-decomp-2}}). The dotted parabola is a lower limit
for binding three bosons in this central field, as deduced from
(\protect{\ref{H3vsH2-2}}).}
\end{figure}
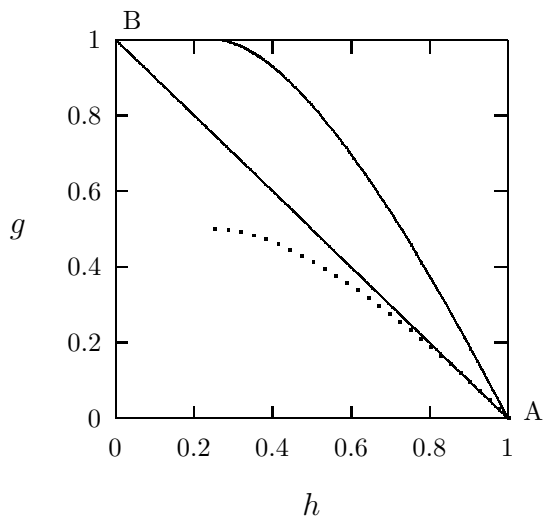


\begin{figure}
\vglue .5cm
\begin{center}
\setlength{\unitlength}{0.240900pt}
\ifx\plotpoint\undefined\newsavebox{\plotpoint}\fi
\sbox{\plotpoint}{\rule[-0.200pt]{0.400pt}{0.400pt}}%
\begin{picture}(900,731)(0,0)
\font\gnuplot=cmr10 at 10pt
\gnuplot
\sbox{\plotpoint}{\rule[-0.200pt]{0.400pt}{0.400pt}}%
\put(220.0,113.0){\rule[-0.200pt]{148.394pt}{0.400pt}}
\put(220.0,113.0){\rule[-0.200pt]{0.400pt}{143.335pt}}
\put(220.0,113.0){\rule[-0.200pt]{4.818pt}{0.400pt}}
\put(198,113){\makebox(0,0)[r]{0}}
\put(816.0,113.0){\rule[-0.200pt]{4.818pt}{0.400pt}}
\put(220.0,232.0){\rule[-0.200pt]{4.818pt}{0.400pt}}
\put(198,232){\makebox(0,0)[r]{0.2}}
\put(816.0,232.0){\rule[-0.200pt]{4.818pt}{0.400pt}}
\put(220.0,351.0){\rule[-0.200pt]{4.818pt}{0.400pt}}
\put(198,351){\makebox(0,0)[r]{0.4}}
\put(816.0,351.0){\rule[-0.200pt]{4.818pt}{0.400pt}}
\put(220.0,470.0){\rule[-0.200pt]{4.818pt}{0.400pt}}
\put(198,470){\makebox(0,0)[r]{0.6}}
\put(816.0,470.0){\rule[-0.200pt]{4.818pt}{0.400pt}}
\put(220.0,589.0){\rule[-0.200pt]{4.818pt}{0.400pt}}
\put(198,589){\makebox(0,0)[r]{0.8}}
\put(816.0,589.0){\rule[-0.200pt]{4.818pt}{0.400pt}}
\put(220.0,708.0){\rule[-0.200pt]{4.818pt}{0.400pt}}
\put(198,708){\makebox(0,0)[r]{1}}
\put(816.0,708.0){\rule[-0.200pt]{4.818pt}{0.400pt}}
\put(220.0,113.0){\rule[-0.200pt]{0.400pt}{4.818pt}}
\put(220,68){\makebox(0,0){0}}
\put(220.0,688.0){\rule[-0.200pt]{0.400pt}{4.818pt}}
\put(343.0,113.0){\rule[-0.200pt]{0.400pt}{4.818pt}}
\put(343,68){\makebox(0,0){0.2}}
\put(343.0,688.0){\rule[-0.200pt]{0.400pt}{4.818pt}}
\put(466.0,113.0){\rule[-0.200pt]{0.400pt}{4.818pt}}
\put(466,68){\makebox(0,0){0.4}}
\put(466.0,688.0){\rule[-0.200pt]{0.400pt}{4.818pt}}
\put(590.0,113.0){\rule[-0.200pt]{0.400pt}{4.818pt}}
\put(590,68){\makebox(0,0){0.6}}
\put(590.0,688.0){\rule[-0.200pt]{0.400pt}{4.818pt}}
\put(713.0,113.0){\rule[-0.200pt]{0.400pt}{4.818pt}}
\put(713,68){\makebox(0,0){0.8}}
\put(713.0,688.0){\rule[-0.200pt]{0.400pt}{4.818pt}}
\put(836.0,113.0){\rule[-0.200pt]{0.400pt}{4.818pt}}
\put(836,68){\makebox(0,0){1}}
\put(836.0,688.0){\rule[-0.200pt]{0.400pt}{4.818pt}}
\put(220.0,113.0){\rule[-0.200pt]{148.394pt}{0.400pt}}
\put(836.0,113.0){\rule[-0.200pt]{0.400pt}{143.335pt}}
\put(220.0,708.0){\rule[-0.200pt]{148.394pt}{0.400pt}}
\put(67,410){\makebox(0,0){$g_{34}$}}
\put(528,-22){\makebox(0,0){$g_{12}$}}
\put(220.0,113.0){\rule[-0.200pt]{0.400pt}{143.335pt}}
\put(724.67,113){\rule{0.400pt}{0.723pt}}
\multiput(725.17,113.00)(-1.000,1.500){2}{\rule{0.400pt}{0.361pt}}
\multiput(723.93,116.00)(-0.477,2.269){7}{\rule{0.115pt}{1.780pt}}
\multiput(724.17,116.00)(-5.000,17.306){2}{\rule{0.400pt}{0.890pt}}
\multiput(718.93,137.00)(-0.482,1.756){9}{\rule{0.116pt}{1.433pt}}
\multiput(719.17,137.00)(-6.000,17.025){2}{\rule{0.400pt}{0.717pt}}
\multiput(712.93,157.00)(-0.477,2.157){7}{\rule{0.115pt}{1.700pt}}
\multiput(713.17,157.00)(-5.000,16.472){2}{\rule{0.400pt}{0.850pt}}
\multiput(707.93,177.00)(-0.482,1.666){9}{\rule{0.116pt}{1.367pt}}
\multiput(708.17,177.00)(-6.000,16.163){2}{\rule{0.400pt}{0.683pt}}
\multiput(701.93,196.00)(-0.482,1.756){9}{\rule{0.116pt}{1.433pt}}
\multiput(702.17,196.00)(-6.000,17.025){2}{\rule{0.400pt}{0.717pt}}
\multiput(695.93,216.00)(-0.482,1.575){9}{\rule{0.116pt}{1.300pt}}
\multiput(696.17,216.00)(-6.000,15.302){2}{\rule{0.400pt}{0.650pt}}
\multiput(689.93,234.00)(-0.482,1.666){9}{\rule{0.116pt}{1.367pt}}
\multiput(690.17,234.00)(-6.000,16.163){2}{\rule{0.400pt}{0.683pt}}
\multiput(683.93,253.00)(-0.482,1.575){9}{\rule{0.116pt}{1.300pt}}
\multiput(684.17,253.00)(-6.000,15.302){2}{\rule{0.400pt}{0.650pt}}
\multiput(677.93,271.00)(-0.482,1.485){9}{\rule{0.116pt}{1.233pt}}
\multiput(678.17,271.00)(-6.000,14.440){2}{\rule{0.400pt}{0.617pt}}
\multiput(671.93,288.00)(-0.485,1.332){11}{\rule{0.117pt}{1.129pt}}
\multiput(672.17,288.00)(-7.000,15.658){2}{\rule{0.400pt}{0.564pt}}
\multiput(664.93,306.00)(-0.485,1.255){11}{\rule{0.117pt}{1.071pt}}
\multiput(665.17,306.00)(-7.000,14.776){2}{\rule{0.400pt}{0.536pt}}
\multiput(657.93,323.00)(-0.482,1.395){9}{\rule{0.116pt}{1.167pt}}
\multiput(658.17,323.00)(-6.000,13.579){2}{\rule{0.400pt}{0.583pt}}
\multiput(651.93,339.00)(-0.485,1.179){11}{\rule{0.117pt}{1.014pt}}
\multiput(652.17,339.00)(-7.000,13.895){2}{\rule{0.400pt}{0.507pt}}
\multiput(644.93,355.00)(-0.485,1.179){11}{\rule{0.117pt}{1.014pt}}
\multiput(645.17,355.00)(-7.000,13.895){2}{\rule{0.400pt}{0.507pt}}
\multiput(637.93,371.00)(-0.485,1.179){11}{\rule{0.117pt}{1.014pt}}
\multiput(638.17,371.00)(-7.000,13.895){2}{\rule{0.400pt}{0.507pt}}
\multiput(630.93,387.00)(-0.485,1.103){11}{\rule{0.117pt}{0.957pt}}
\multiput(631.17,387.00)(-7.000,13.013){2}{\rule{0.400pt}{0.479pt}}
\multiput(623.93,402.00)(-0.488,0.956){13}{\rule{0.117pt}{0.850pt}}
\multiput(624.17,402.00)(-8.000,13.236){2}{\rule{0.400pt}{0.425pt}}
\multiput(615.93,417.00)(-0.485,1.026){11}{\rule{0.117pt}{0.900pt}}
\multiput(616.17,417.00)(-7.000,12.132){2}{\rule{0.400pt}{0.450pt}}
\multiput(608.93,431.00)(-0.488,0.890){13}{\rule{0.117pt}{0.800pt}}
\multiput(609.17,431.00)(-8.000,12.340){2}{\rule{0.400pt}{0.400pt}}
\multiput(600.93,445.00)(-0.488,0.890){13}{\rule{0.117pt}{0.800pt}}
\multiput(601.17,445.00)(-8.000,12.340){2}{\rule{0.400pt}{0.400pt}}
\multiput(592.93,459.00)(-0.485,0.950){11}{\rule{0.117pt}{0.843pt}}
\multiput(593.17,459.00)(-7.000,11.251){2}{\rule{0.400pt}{0.421pt}}
\multiput(585.93,472.00)(-0.488,0.824){13}{\rule{0.117pt}{0.750pt}}
\multiput(586.17,472.00)(-8.000,11.443){2}{\rule{0.400pt}{0.375pt}}
\multiput(577.93,485.00)(-0.489,0.669){15}{\rule{0.118pt}{0.633pt}}
\multiput(578.17,485.00)(-9.000,10.685){2}{\rule{0.400pt}{0.317pt}}
\multiput(568.93,497.00)(-0.488,0.824){13}{\rule{0.117pt}{0.750pt}}
\multiput(569.17,497.00)(-8.000,11.443){2}{\rule{0.400pt}{0.375pt}}
\multiput(560.93,510.00)(-0.488,0.758){13}{\rule{0.117pt}{0.700pt}}
\multiput(561.17,510.00)(-8.000,10.547){2}{\rule{0.400pt}{0.350pt}}
\multiput(552.93,522.00)(-0.489,0.611){15}{\rule{0.118pt}{0.589pt}}
\multiput(553.17,522.00)(-9.000,9.778){2}{\rule{0.400pt}{0.294pt}}
\multiput(543.93,533.00)(-0.488,0.692){13}{\rule{0.117pt}{0.650pt}}
\multiput(544.17,533.00)(-8.000,9.651){2}{\rule{0.400pt}{0.325pt}}
\multiput(535.93,544.00)(-0.489,0.611){15}{\rule{0.118pt}{0.589pt}}
\multiput(536.17,544.00)(-9.000,9.778){2}{\rule{0.400pt}{0.294pt}}
\multiput(526.93,555.00)(-0.489,0.553){15}{\rule{0.118pt}{0.544pt}}
\multiput(527.17,555.00)(-9.000,8.870){2}{\rule{0.400pt}{0.272pt}}
\multiput(517.93,565.00)(-0.489,0.553){15}{\rule{0.118pt}{0.544pt}}
\multiput(518.17,565.00)(-9.000,8.870){2}{\rule{0.400pt}{0.272pt}}
\multiput(508.93,575.00)(-0.489,0.553){15}{\rule{0.118pt}{0.544pt}}
\multiput(509.17,575.00)(-9.000,8.870){2}{\rule{0.400pt}{0.272pt}}
\multiput(498.92,585.59)(-0.495,0.489){15}{\rule{0.500pt}{0.118pt}}
\multiput(499.96,584.17)(-7.962,9.000){2}{\rule{0.250pt}{0.400pt}}
\multiput(489.74,594.59)(-0.553,0.489){15}{\rule{0.544pt}{0.118pt}}
\multiput(490.87,593.17)(-8.870,9.000){2}{\rule{0.272pt}{0.400pt}}
\multiput(479.92,603.59)(-0.495,0.489){15}{\rule{0.500pt}{0.118pt}}
\multiput(480.96,602.17)(-7.962,9.000){2}{\rule{0.250pt}{0.400pt}}
\multiput(470.51,612.59)(-0.626,0.488){13}{\rule{0.600pt}{0.117pt}}
\multiput(471.75,611.17)(-8.755,8.000){2}{\rule{0.300pt}{0.400pt}}
\multiput(460.72,620.59)(-0.560,0.488){13}{\rule{0.550pt}{0.117pt}}
\multiput(461.86,619.17)(-7.858,8.000){2}{\rule{0.275pt}{0.400pt}}
\multiput(451.21,628.59)(-0.721,0.485){11}{\rule{0.671pt}{0.117pt}}
\multiput(452.61,627.17)(-8.606,7.000){2}{\rule{0.336pt}{0.400pt}}
\multiput(441.21,635.59)(-0.721,0.485){11}{\rule{0.671pt}{0.117pt}}
\multiput(442.61,634.17)(-8.606,7.000){2}{\rule{0.336pt}{0.400pt}}
\multiput(431.21,642.59)(-0.721,0.485){11}{\rule{0.671pt}{0.117pt}}
\multiput(432.61,641.17)(-8.606,7.000){2}{\rule{0.336pt}{0.400pt}}
\multiput(420.54,649.59)(-0.943,0.482){9}{\rule{0.833pt}{0.116pt}}
\multiput(422.27,648.17)(-9.270,6.000){2}{\rule{0.417pt}{0.400pt}}
\multiput(409.82,655.59)(-0.852,0.482){9}{\rule{0.767pt}{0.116pt}}
\multiput(411.41,654.17)(-8.409,6.000){2}{\rule{0.383pt}{0.400pt}}
\multiput(399.82,661.59)(-0.852,0.482){9}{\rule{0.767pt}{0.116pt}}
\multiput(401.41,660.17)(-8.409,6.000){2}{\rule{0.383pt}{0.400pt}}
\multiput(388.93,667.59)(-1.155,0.477){7}{\rule{0.980pt}{0.115pt}}
\multiput(390.97,666.17)(-8.966,5.000){2}{\rule{0.490pt}{0.400pt}}
\multiput(377.93,672.59)(-1.155,0.477){7}{\rule{0.980pt}{0.115pt}}
\multiput(379.97,671.17)(-8.966,5.000){2}{\rule{0.490pt}{0.400pt}}
\multiput(366.93,677.59)(-1.155,0.477){7}{\rule{0.980pt}{0.115pt}}
\multiput(368.97,676.17)(-8.966,5.000){2}{\rule{0.490pt}{0.400pt}}
\multiput(355.02,682.60)(-1.505,0.468){5}{\rule{1.200pt}{0.113pt}}
\multiput(357.51,681.17)(-8.509,4.000){2}{\rule{0.600pt}{0.400pt}}
\multiput(344.02,686.60)(-1.505,0.468){5}{\rule{1.200pt}{0.113pt}}
\multiput(346.51,685.17)(-8.509,4.000){2}{\rule{0.600pt}{0.400pt}}
\multiput(331.50,690.61)(-2.248,0.447){3}{\rule{1.567pt}{0.108pt}}
\multiput(334.75,689.17)(-7.748,3.000){2}{\rule{0.783pt}{0.400pt}}
\multiput(320.50,693.61)(-2.248,0.447){3}{\rule{1.567pt}{0.108pt}}
\multiput(323.75,692.17)(-7.748,3.000){2}{\rule{0.783pt}{0.400pt}}
\multiput(308.94,696.61)(-2.472,0.447){3}{\rule{1.700pt}{0.108pt}}
\multiput(312.47,695.17)(-8.472,3.000){2}{\rule{0.850pt}{0.400pt}}
\put(292,699.17){\rule{2.500pt}{0.400pt}}
\multiput(298.81,698.17)(-6.811,2.000){2}{\rule{1.250pt}{0.400pt}}
\put(281,701.17){\rule{2.300pt}{0.400pt}}
\multiput(287.23,700.17)(-6.226,2.000){2}{\rule{1.150pt}{0.400pt}}
\put(269,703.17){\rule{2.500pt}{0.400pt}}
\multiput(275.81,702.17)(-6.811,2.000){2}{\rule{1.250pt}{0.400pt}}
\put(257,704.67){\rule{2.891pt}{0.400pt}}
\multiput(263.00,704.17)(-6.000,1.000){2}{\rule{1.445pt}{0.400pt}}
\put(245,705.67){\rule{2.891pt}{0.400pt}}
\multiput(251.00,705.17)(-6.000,1.000){2}{\rule{1.445pt}{0.400pt}}
\put(232,706.67){\rule{3.132pt}{0.400pt}}
\multiput(238.50,706.17)(-6.500,1.000){2}{\rule{1.566pt}{0.400pt}}
\put(220.0,708.0){\rule[-0.200pt]{2.891pt}{0.400pt}}
\end{picture}
\end{center}
\vglue .5cm
\caption{\label{Fig2}Lower limit for binding four particles with
masses $(m,m,M,M)$.
The couplings $g_{12}$ and $g_{34}$ are normalized to the critical
coupling
for binding $(m,m)$ and $(M,M)$, respectively. Meanwhile the $(m,M)$
coupling
should be at least half the critical coupling for binding $m$ to $M$.
A value $M/m=2$ is assumed here for the drawing.}
\end{figure}
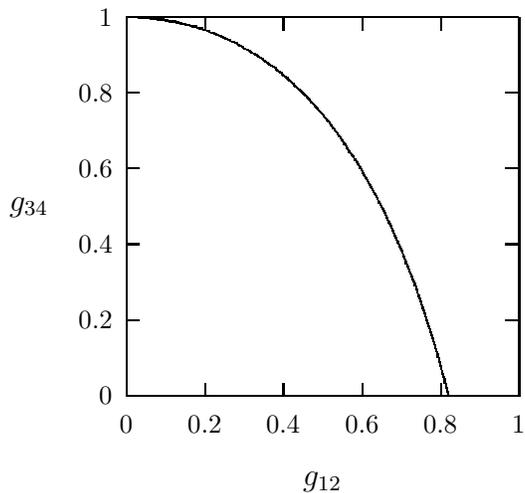


\begin{thebibliography}{10}

\bibitem{Bang}
M.V. Zhukov, B.V. Danilin, D.V. Fedorov, J.M. Bang, I.S. Thompson,
and J.S.
  Vaagen, Phys. Rep. {\bf 231} (1993) 151.

\bibitem{Jensen3b}
D.V. Fedorov, A.S. Jensen and K. Riisager, Phys. Rev. {\bf C49}
(1994) 201.

\bibitem{Chinois}
Zhougzhou Ren, Phys. Rev. {\bf C49} (1994) 1281; A. Cs{\'o}t{\'o},
Phys. Rev.
  {\bf C48}(1993) 165; K. Varga, Y. Suzuki and R.G. Lowas, Nucl.
Phys. {\bf
A571}
(1994)  447.

\bibitem{Efimov}
V. Efimov, Phys. Lett. {\bf 33B} (1970) 560; Sov. J. Nucl. Phys. {\bf
12}
  (1971) 581; S. Albeverio, R. H{\o}egh-Krohn and T.T. Wu, Phys.
Lett. {\bf
  83A} (1981) 101; D.V. Fedorov and A.S. Jensen, Phys. Rev. Lett.
{\bf 71}
  (1993) 4103.

\bibitem{Thomas}
L.H. Thomas, Phys. Rev. {\bf 47} (1935) 903; A. Delfino, K. Adhikari,
L. Tomio
  and T. Frederico, Phys. Rev. {\bf C46} (1992) 471.

\bibitem{Jackson}
J.M. Blatt and J.D. Jackson, Phys. Rev. {\bf 76} (1949) 18.

\bibitem{Dyson}
F.J. Dyson and A. Lenard, J. Math. Phys. {\bf 8} (1967) 621.

\bibitem{Gillepsie}
J.J. Benayoun, C. Gignoux, J. Gillepsie, and A. Laverne, Lettere al
Nuovo
  Cimento {\bf 8} (1973) 414.

\bibitem{Post}
R.L.~Hall and H.R.~Post, Proc.~Phys.~Soc.\ {\bf 90} (1967) 381.

\bibitem{BMR1}
J.-L.~Basdevant, A.~Martin and J.-M.~Richard, Nucl.~Phys.\ {\bf B343}
(1990)
  60.

\bibitem{BMR2}
J.-L.~Basdevant, A.~Martin and J.-M.~Richard, Nucl.~Phys.\ {\bf B343}
(1990)
  69.

\bibitem{BMR3}
J.-L.~Basdevant, A.~Martin, J.-M.~Richard and T.T. Wu, Nucl.~Phys.\
{\bf B393}
  (1993) 111.

\bibitem{quantumc}
See, for instance, M. Kaminura, Phys. Rev. {\bf A38} (1988) 621; D.B.
Kinghorn
  and R.D. Poshusta, Phys. Rev. {\bf A47}(1993) 3671; and references
therein.

\bibitem{Thir}
See, e.g., W.~Thirring, A Course in Mathematical Physics, {\bf
Vol.~3}: {\it
  Quantum Mechanics of Atoms and Molecules} (Springer Verlag, New
York, 1981).

\bibitem{Gal}
For \protect{$S=-2$} hypernuclei, see, for instance, A. Gal, in {\it
Flavour
  and Spin in Hadronic and Electromagnetic Interactions}, Proc. Turin
Workshop,
  sept. 1992, ed. F. Balestra, R. Bertini, and R. Garfagnini (Italian
Physical
  Society, Bologna, 1993); see also several contributions in {\it
Hypernuclear
  and Strange-Particle Physics}, Proc. Int. Symp., Shimoda, Japan,
dec. 1991,
  ed. T. Fukuda et al., Nucl. Phys. {\bf A547} (1992).

\end{thebibliography}
\end {document}